\documentclass[12pt]{article}
\usepackage[a4paper, margin=1.22in]{geometry}
\usepackage{color}
\usepackage{psfrag}
\usepackage{enumerate}
\usepackage{amsmath,amssymb,calc,amsfonts}
\usepackage{latexsym}
\def\beq{\begin{eqnarray}}
\def\eeq{\end{eqnarray}}

\def\={\stackrel{\Delta}{=}}

\def\lie{\pounds}

\usepackage[utf8]{inputenc}

\title{Joining Spacetimes on Fractal Hypersurfaces}
\author{Ayan Chatterjee\footnote{ayan.theory@gmail.com} ~and~ Ankit Anand\footnote{ankitanandp94@gmail.com}\\
Department of Physics and Astronomical Science\\
Central University of Himachal Pradesh, Dharamshala-176206, India.}

\usepackage{graphicx}
\begin{document}
\date{}
\maketitle

\begin{abstract}
The theory of fractional calculus is attracting a lot of attention
from mathematicians as well as physicists. The fractional generalisation of the well- known ordinary calculus is being 
used extensively in many fields, particularly in understanding stochastic process and fractal dynamics. In this paper, we apply 
the techniques of fractional calculus to study some specific modifications of the geometry of submanifolds. Our generalisation is applied to extend the Israel formalism which is used to 
glue together two spacetimes across a timelike, spacelike or a null hypersurface. In this context, we show that the fractional extrapolation leads to some 
striking new results. More precisely we demonstrate that, in contrast to the original Israel formalism, where many spacetimes can only 
be joined together through an intermediate thin hypersurface of matter satisfying some non- standard energy conditions, the fractional 
generalisation allows these spacetimes to be smoothly sewed together without any such requirements on the stress tensor of
the matter fields. We discuss 
the ramifications of these results for spacetime structure and the possible implications for
gravitational physics.    

\end{abstract}

\section{Introduction}
The theory of fractional calculus has been considered a classical but obscure corner of mathematics
\cite{oldham_spanier, miller_ross, samco_kilbas_marchev}. It remained, until a few decades, a field by mathematicians, for mathematicians and
of purely theoretical interest. Though it played a crucial role in the development of 
Abel's theory of integral equations and many mathematicians like Liouville, 
Riemann, Heaviside and Hilbert took an active interest in it, fractional calculus found limited applications 
and was referred to only occasionally, to simplify complicated solutions. For example, this formalism has been used quite often
 to simplify the solutions of both the diffusion as well as the wave equation (for example, see \cite{courant_Hilbert}, and \cite{widder}). 

During the last few decades, however, this theory has found important applications for large number of practical 
real life situations. Indeed, fractional calculus is providing excellent tools to develop models 
of polymers and materials \cite{douglas1, douglas2}. In particular, it has been found that to understand  properties of various materials which require long- range order to hold, fractional calculus provides a sound platform \cite{Koeller}. Fractional calculus have also been found to naturally incorporate some subtle effects in 
the dynamics of fluids, and these have found important applications in understanding mechanical, chemical and electrical
properties of nano- fluids. However, possibly the most prominent application of 
these derivatives of non- integer order has been in the theory of fractals \cite{Mandelbrot}. It has been found that for many stochastic processes, the phenomena progresses through increments which are not independent, but instead tend to retain some memory of previous increment, though not necessarily the immediately previous increment \cite{Mandelbrot,Beran, Giona_Roman, Roman_Giona, west}. In other words, these are random processes with long term memory. The theory of fractional Brownian motion, which 
provides a very natural explanation for these effects, incorporates these persistence effects (or anti- persistence effects) though the fractional modification of the usual Brownian formula relating displacement and time \cite{Mandelbrot_ness}. It has now been understood
that statistically speaking, all the naturally occurring signals are of the Weierstrass type, \emph{i.e} continuous but non- differentiable \cite{Mandelbrot,burrough} (here differentiation is in the sense of the usual calculus) and indeed, such Weierstrass- like functions arise even in many quantum mechanical situations.
For example, it has been shown that many quantum mechanical problems involving discontinuous potentials possesses energy 
spectrum of the Weierstrass type \cite{Berry_Lewis}. Furthermore, the  Feynman paths in the path integral formulation of quantum mechanics are
also examples of these kind \cite{Feynman}.  However, the most significant discovery has been that, though the naturally occuring functions are of the Weierstrass type and are endowed with a fractal dimension, \emph{they are fractionally differentiable} and that the maximal order of differentiability is related to the box- dimension of the function \cite{Kolwankar_Gangal, Edgar}. Thus, fractional calculus has been highly advantageous in modelling dynamical processes in self- similar systems and for analysing processes which generate chaotic signals and are apparently irregular.  

In this paper, we apply the techniques of fractional calculus 
to general relativity. As is well known, the issue of final state of gravitational collapse
is a long standing open question in general relativity. The appearance of spacetime singularity
reveals the domain of failure of the classical theory of general relativity \cite{wald}. Quite naturally,
it is assumed that general relativity must be corrected to eliminate these failures. Both the string theories and effective field theories 
necessitate that one must add terms involving higher order as well as higher derivatives in the Riemann tensor to incorporate 
the effects of physics at small scales. It is a general hope that these higher order corrections will 
certainly get rid of the singularities \cite{Ortin}. However, in absence
of any comprehensive proof of these expectations, we propose to look at another alternative which may present itself
at the small scales. As we shall see in the subsequent sections,
fractional calculus, in any of it's possible alternative forms, define differentiation through an integration. Hence, it naturally 
incorporates non- local spacetime correlations and long- range interactions, which are expected to be
natural at high energy scales, into account. Thus, many 
subtle non- local effects may manifest itself if one replaces the ordinary differentiation by it's fractional counterpart. 
One may immediately ask as to where should one look for such non- local terms to arise physically and envisage
the regions of strong gravity where the classical theory of general relativity is known to require modifications\footnote{Most of these terms contribute non- local effects into the Green's function. It should not be surprising 
if many of the effects of the fractional generalisation arise naturally in the string theories or any other quantum theory.}.
An obvious candidate for the strong gravity regime is the black hole region since black holes are created due to gravitational
collapse of matter fields in an intense  gravitational field. There are two regions of the 
black hole spacetime which are ideally suited for the fractional effects to manifest itself. First is the horizon, which for small mass 
black holes are regions of intense gravitional field and second, near the singularity where the effects of strong gravity, though invisible to the asymptotic observer, are most spectacular. In either of these two situations, possibilities of long range order have
quite interesting repercussions on modelling of spacetime.

Let us first discuss the region near the horizon. The black hole horizon, here taken to be an event horizon, is a null expansion- free 
hypersurface which lies in the region adjoining two spacetimes. So, the
horizon may be thought of as a null hypersurface which glues the two spacetimes. 
Naturally, the joining of two spacetimes through the hypersurface requires that some conditions on the spacetime variables be satisfied
on the hypersurface. The Israel- Darmois- Lanczos (IDL) junction condition demands that the metric on
either side of the horizon, when pulled back to
the hypersurface, must be continuous \cite{poisson}. In contrast, the extrinsic curvature of the horizon is not required to be continuous. 
In fact, consistency requires that the Riemann tensor and hence the extrinsic cuvature on the hypersurface admit delta function 
singularities. Using the Einstein equations, the Ricci part of this singularity is related to
the stress tensor. Thus, the IDL condition only requires that the difference of the extrinsic curvatures of 
the hypersurface as embedded in these two spacetimes, must be proportional to the stress- energy tensor living on 
the horizon. In other words, due to the geometry itself, the hypersurface comes naturally equipped with a energy- momentum tensor.
These junction conditions on the horizon has given rise to speculations of constructing a singularity free spacetimes,
and in particular, non- singular black hole interiors, in the
following way \cite{Shen, Barrabes_Israel}: 
Take the exterior of the Schwarzschild horizon as the future spacetime region and the interior to de- Sitter horizon
as the past spacetime region. The boundary between these two regions, the common hypersurface to these
two regions, will be a thin null hypersurface endowed with some specific energy- momentum tensor derived from the 
IDL matching conditions. Thus, one may have well defined matching conditions to create a singularity
free universe, with the exterior a Schwarzschild spacetime while the interior being a de- Sitter spacetime. However, in most of the cases,
the matching conditions leads to energy- momentum tensors which
violate some of the well known energy conditions.  In \cite{Barrabes_Israel}, there
have been attempts at constructing a singularity free universe by adjoining the de- Sitter interior with 
the inner horizon of a Reissner- Nordstrom black hole with a particular values of charge and mass. In 
this particular case, the matching is smooth, with no requirement of
any energy momentum tensor. In general situations, 
these matchings are not smooth
and require energy condition violating energy- momentum tensors on the matching hypersurface.

The second point is related to the another such attempt where, the de- Sitter spacetime is glued to the Schwarzschild interior
though a spacelike hypersurface. This attempt was made by \cite{Frolov_markov}, in their famous proposal of \emph{limiting curvature}. 
They devised
a model in which the Schwarzschild metric inside the black hole region is matched to a de- Sitter one at some spacelike
junction surface which represent a thin transition layer. As a requirement of their proposal, this layer is placed at a region
very close to the singularity where the curvature reaches it's limiting value. However again, for general singularity free matchings
of the above kind, the junction layers admit energy momentum tensors which violate energy conditions. In particular, the
effective stress- energy tensor of the model \cite{Frolov_markov} violates the weak energy condition. In fact, in almost
all similar attempts of creating singularity free models like that of \cite{Frolov_markov}, have energy condition violations.
These kind of violations are actually characteristic of quantum effects which become important in strong gravitational field. 

As a remedy to these energy condition violations, we argue in this paper that the notion of fractional derivatives offers a possibility of creating singularity free universe through smooth matching of spacetimes. More precisely, we demonstrate the following: First, that the IDL 
junction conditions both for timelike/spacelike as well as for null hypersurfaces
are modified due to the fractional generalisation of the spacetime connection. This fractional generalisation of the IDL conditions
will in turn modify the energy-momentum tensor on the hypersurface\footnote{The fractional generalisation developed in
this paper assumes that fractal like structures are present at very high energy scales, just as they are present at low energy scales. There is no
experimental basis for such assumptions. However, this assumption leads to
some interesting consequesnces as developed in this paper.}. Secondly, using specific examples,
we show that this generalisation allows us to fix the conditions on the junction shell 
in such a way that the Schwarzschild or the Reissner- Nordstrom spacetimes can always be smoothly matched 
to the de- Sitter spacetime in the interior without any energy condition violating requirement
on the energy- momentum tensors of the adjoining shell.

The paper is organised as follows: In the next section, section 2,  we briefly discuss the mathematical formalism of fractional calculus and 
the relevant notations. In sections 3 and 4, we introduce 
the notations for timelike/spacelike and null hypersurfaces and discuss the generalisations of the IDL junction conditions 
for fractional exponents. We also argue that these generalised junction conditions leads to smooth joining 
of spacetimes, which otherwise are known to be joined only through a thin shell of matter. The implications of 
these are discussed in the Discussion section.  

\section{Mathematical preliminaries}
Let us discuss some notations useful for the mathematical formulation of geometry of hypersurfaces.
Let us consider a $4$ dimensional spacetime $(\mathcal M,{g})$ with signature ($-,+,+,+$). Let a hypersurface $\Delta$ 
be embedded in $\mathcal M$ and is given by $f:\Delta\rightarrow \mathcal M$. We shall assume that
the embedding relation is such that the restriction of $f$ to the image of $\Delta$ is $C^{\infty}$.
Let, $\{x^\mu\}$ be a local coordinate chart  on $\mathcal M$ and $\{y^a\}$ be a local coordinate chart 
on $\Delta$. The embedding relation implies $x^{\mu}=x^{\mu}(y^a)$. Let, $g_{\mu\nu}$ be the metric on the spacetime in terms of it's local coordinates.
The first fundamental form or the induced metric on $\Delta$ is the pull back of the metric $g$ under the map $f$. 
In the local coordinates this can be written as $h_{ab}$. 
\begin{equation}
h_{ab}\equiv g(\partial_a,\partial_b)=\frac{\partial x^\mu}{\partial y^a}\frac{\partial x^\nu}{\partial y^b}\, g(\partial_\mu,\partial_\nu)
=e^{\mu}_{a}e^{\nu}_{b}\, g_{\mu\nu},
\end{equation}
where,  $\left(\partial x^\mu/\partial y^a\right)=e^{\mu}_{a}$ and we have used that $e^{\mu}_{a}\,\partial_\mu$ is 
the push forward of the purely tangential vector 
field $\partial_a$ onto the full spacetime $\mathcal M$. 
One may define a linear connection and hence a derivative operator on the spacetime.
Let, $T\mathcal M$ denote the tangent bundle on $\mathcal M$ and let, $X$ and $Y$ 
are two arbitrary vector fields on it. The covariant derivative is a linear map 
\begin{eqnarray}
&\nabla: &T\mathcal M \otimes T\mathcal M\rightarrow T\mathcal M\\
&& (X,Y)\rightarrow \nabla_X Y.
\end{eqnarray}
The Riemannian theory assumes the covariant derivative to be metric compatible, $\nabla_{Z}\,g(X,Y)=g\,(\nabla_{Z}X,Y)+ g\,(X,\nabla_{Z}Y)$. 
On the tangent bundle one may also define a covariant derivative ($D$) on 
$\Delta$ using the Gauss decomposition formula: For 
$X,Y\in\Delta$,
\begin{eqnarray}\label{Gauss-decompo}
&D:& T\Delta \otimes T\Delta \rightarrow T\Delta\\
&&\nabla_XY = D_XY+K(X,Y).
\end{eqnarray}
$D_XY$ is purely tangential and $K(X,Y)$ is an element of the normal bundle and refereed to as the extrinsic curvature. 
The Gauss equation also implies along with that metric compatibility of $\nabla$ with $g$ 
that the derivative operator $D_{a}$ is metric compatibile with the metric
on the hypersurface $h_{ab}$ (\emph{i.e.} $D_{a}h_{bc}=0$).
In terms of the local coordinate charts, the Gauss equation gives the following expression for the derivative operator (for $X\equiv \partial_{a}$):
\begin{equation}
D_{a}Y_{b}=e^{\mu}_{a}\,e^{\nu}_{b}\,\nabla_{\mu}Y_{\nu}.
\end{equation}
The extrinsic curvature can also be defined for the hypersurface in terms of the local coordinates. The normal bundle for the hypersurface is
one dimensional. Let, $n^{\mu}$ be the normal. The extrinsic curvature is
\begin{equation}
K_{ab}=e^{\mu}_{a}\,e^{\nu}_{b}\,\nabla_{\mu}n_{\nu}=(1/2)\,\left(\lie_{n}g_{\mu\nu}\right)\,e^{\mu}_{a}\,e^{\nu}_{b}.
\end{equation}
For our later use, let us give the Gauss equation in terms of the local coordinates:
\begin{equation}
R_{\mu\nu\lambda\sigma}\,e_{a}^{\mu}\,e_{b}^{\nu}\,e_{c}^{\lambda}\,e_{d}^{\sigma}=R_{abcd}-\left(K_{ad}K_{bc}-K_{ac}K_{bd}\right).
\end{equation}
The Codazzi equation in local coordinates is given by:
\begin{equation}
R_{\mu\nu\lambda\sigma}\,n^{\mu}\,e_{b}^{\nu}\,e_{c}^{\lambda}\,e_{d}^{\sigma}=K_{ab|c}-K_{ac|b},
\end{equation}
where $|$ denotes the covariant derivative with respect to the coordinates on the hypersurface.

Several of these spacetime functions have different values on either sides of a hypersurface. Then, it is required to express their continuity across
the hypersurface. A useful and prominent example of this idea is
that of the Israel- Darmois- Lanczos (IDL) junction condition \cite{poisson}. 
Consider a hypersurface $\Delta$ which partitions the spacetime into two regions $(M_{+}, g_{+})$
with coordinates $\{x_{+}^{\mu}\}$ and $(M_{-}, g_{-})$ with coordinates $\{x_{-}^{\mu}\}$. The spacetime 
$M_{+}$ is assumed to be to the future of the spacetime $M_{-}$. Quite naturally, it is not generally 
true that the metrics on these two spacetimes could be continuously matched across the hypersurface $\Delta$ 
(The either side of the hypersurface $\Delta$ has been installed with coordinates $\{y^{a}\}$). The 
discontinuity in the metric would be reflected in the fact that Riemann tensor would have a delta- function
singularity on the hypersurface. The Israel junction conditions provides a method to smoothly match these hypersurfaces
by using the following trick: relate the Ricci part of the singular Riemann curvature tensor to the surface stress- tensor
using the Einstein equations. 
For spacelike hypersurfaces, the Israel junction conditions for a smooth joining of
hypersurfaces at $\Delta$ is given by 
\begin{equation}
[h_{ab}]=0=[K_{ab}]
\end{equation}
where $[A]\equiv A(M^{+})|_{\Delta}-A(M^{-})|_{\Delta}$. However, if the extrinsic curvature is
not identical on both the sides on the hypersurface $\Delta$, the surface stress tensor ($S_{ab}$) on the hypersurface is 
\begin{equation}\label{emtensor_eqn}
8\pi\,S_{ab}=[K_{ab}]-[K]h_{ab}.
\end{equation}
However, on the null surface, the standard extrinsic curvature corresponding the normal of the hypersurface (which is also the tangent 
to null hypersurface) is always continuous and hence, one needs to define a transverse curvature \cite{poisson}. The 
metric induced on the null hypersurface is again continuous but the discontinuities in the components of the transverse curvature
is related to the energy- momentum tensor induced on the this hypersurface.

In deriving the above relations, we have implicitly made two crucial assumtions: First, that the point functions are continuous and differentiable
in the region under consideration. However, it may happen the scalar, vector or the tensor functions are only fractionally differentiable. In that case, the limiting values defined
by our ordinary differential calculus become singular on the hypersurface.
Thus, in addition to the IDL conditions, their fractional character must also be taken into account. Secondly, the spacetime connection is assumed to be 
a Levi- Civita connection. This arises since the spacetime is assumed
to be a Riemannian spacetime and hence, the spacetime metric is compatible with the covariant derivative ($\nabla_{\gamma}\,g_{\alpha\beta}$=0). The Gauss decomposition, eqn.\eqref{Gauss-decompo}, then imples that the connection on the hypersurface is also a Levi- Civita connection and that
the extrinsic curvature is uniquely detremined in terms of
this connection. However, it may happen that in the strong gravity 
regime we are interested in, the spacetime is slightly modified from it's 
Riemannian character and that the connection is not Levi- Civita connection
derived from the metric.
Quite naturally, in such a situation, the Gauss decomposition implies 
that the connection on the hypersurface will also be modified and the expression of the extrinsic curvature will also change.\footnote{In the appendix \ref{appendix3}, we have developed a non- Levi-Civita connection based on a notion of fractional derivative and have shown to lead to modification of the tensor functions and the Einstein equations.}

In \cite{Krisch}, a fractional generalisation of the Lie derivative has been proposed and utilised to generalise the definition of the extrinsic 
curvature for non- null hypersurfaces.  In this fractional generalisation, 
which is based on Caputo's modification of the Riemann- Liouville definition
of fractional derivative  (see the appendix), the usual definition
of the extrinsic curvature $K_{ab}=(1/2)(\lie_{n}g_{\alpha\beta})\,e^{\alpha}_{a}e^{\beta}_{b}$ is modified
to give:
\begin{eqnarray}\label{qlie_expansion}
{}^{q}K_{ab} &=&\frac{1}{2}\left({}^{q}\lie_n g_{\alpha \beta}\right)\, e_a ^\alpha\, e_b ^\beta \nonumber\\
&=&\frac{1}{2}\,\left[n^{\gamma}\, \mathcal{D}^q _{r-\Delta, \gamma}\, g_{\alpha \beta} + 
g_{\gamma \beta} \mathcal{D}^q _{r-\Delta,r} n^\gamma + 
g_{\alpha \gamma} \mathcal{D}^q _{r-\Delta,r} n^\gamma \right]\,e_a ^\alpha\, e_b ^\beta.
\end{eqnarray}
Here, the superscript $q$ denotes the fractional parameter, $0<q\le 1$ (see the appendix) and $\mathcal{D}^q _{r-\Delta,r}$
denotes the derivative:
\begin{equation}
\mathcal{D}^q_{r-\Delta,r}(g_{\alpha\beta})=\frac{\Gamma{(2-q)}}{\Gamma{(1-q)}\,\Delta^{1-q}} \int_{r-\Delta}^{r} 
\frac{\partial g_{\alpha\beta}(w)}{\partial w} (r-w)^{-q} dw,
 \end{equation}
where the integration is carried out from a spacepoint $r-\Delta$ to $r$. In the context of matching of spacetimes across hypersurface, $\Delta$ is taken to be
the thickness of the hypersurface. The junction conditions will be modified from eqn. \eqref{emtensor_eqn} to
\begin{equation}
8\pi\,{}^{q}S_{ab}=[{}^{q}K_{ab}]-[{}^{q}K]h_{ab}.
\end{equation}
Naturally, because of the definition of the derivative, it has a non- local character imbedded 
into it. In the following sections, we shall utilize this generalisation of the definition of extrinsic curvature to
modify the junction conditions for spacelike/timelike as well as null hypersurfaces. Additionally, we shall show that
the junction conditions lead to a smooth matching of hypersurfaces.

\section{Junction conditions for non- null hypersurfaces}
Let us consider a non- null hypersurface $\Delta$. As discussed previously, the junction condition for the
smooth joining of spacetimes along a timelike or spacelike hypersurface $\Delta$ is given by the following two conditions:
$[h_{ab}]=0$ and $[K_{ab}]=0$. On the other hand, for joining spacetimes which contribute non- equal extrinsic curvatures on the hypersurface, 
a thin layer of matter is assumed to exist on the hypersurface with stress tensor $S_{ab}=(\epsilon/8\pi)([K_{ab}]-Kh_{ab})$. 
The quantity $\epsilon=n\cdot n$, distinguishes spacelike hypersurfaces ($\epsilon=-1$) from timelike  ones ($\epsilon=-1$).

As described in the previous subsection, the junction conditions differ if the fractional derivatives are used. For the non- null
hypersurface, we elaborate on this method through two explicit examples. In the first example, we give a detail step by step calculation
showing the matching of a slowly rotating Kerr metric to a Minkowski metric on a timelike hypersurface. We show that depending on the 
width of the shell, the energy momentum tensor of the shell changes. We utilize this observation in the second example, which
deals with matching of a Schwarzschild spacetime with a de- Sitter spacetime on a spacelike hypersurface. Again the energy- momentum 
tensor residing on the thin shell differs substantially from the standard results.

\subsection{Joining Minkowski and slowly rotating Kerr metrics}
Let us consider the metric of a Kerr spacetime in the slow- rotation approximation.  We shall
assume a shell of mass $M$ and angular momentum $J$ in the spacetime. The exterior spacetime ($\mathcal{M}^{+}$)
has the following metric:
\begin{equation}
ds^2=-f(r)\,dt^2+f^{-1}(r)\,dr^2+r^2 \,d\Omega^2 - \frac{4Ma}{r}\sin^2 \theta\, dt\,d\phi,
\end{equation}
where $f(r)=(1-2M/r)$ and $a=(J/M)\ll M$, is a parameter for the angular momentum which is usually used to 
replace the  shell's angular momentum $J$. Let us assume that the shell is located at $r=R_{0}$.
The induced metric on the shell becomes: 
\begin{equation}
ds^2 _\Sigma=-f(R_{0})\, dt^2  + R_{0}^2\, d\Omega^2 
- \frac{4Ma}{R_{0}}\,\sin^2 \theta\, dt\, d\phi.
\end{equation}
Using the definitions, $\psi =(\phi-\omega t)$ with $\omega=(2Ma/r^3)$, and keeping terms upto first order of $a$,
we get the induced metric to be
\begin{equation}
h_{ab}dy^a dy^b=-f(r)\,dt^2 +R_{0}^2\,(d\theta^2 +\sin^2 \theta\, d\psi^2).
\end{equation}
We shall use $y^a=(t,\theta, \psi)$ as the co-ordinates on the shell and the parametric equations for 
the hypersurface in the form $x^\alpha=x^\alpha(y^a)$ are $t=t, \theta=\theta\, \mbox{and}\, \phi=(\psi+\omega t)$. The 
shell's unit normal is $n_\alpha=f(r)^{-1/2}\,\partial_\alpha r$, which in the coordinates is given by:
\begin{eqnarray}
n^\alpha=\left(0,\sqrt{1-2M/r},0,0 \right).
\end{eqnarray}
Now let's calculate the non vanishing components of extrinsic curvature. 
The definition of transverse component of the fractional generalisation is:
\begin{eqnarray}\label{qlie_expansion}
{}^{q}K_{ab} &=&\frac{1}{2}\left({}^{q}\lie_n g_{\alpha \beta}\right)\, e_{a}^\alpha\, e_b ^\beta \nonumber\\
&=&\frac{1}{2}\,\left[n^{\gamma}\, \mathcal{D}^q _{r-\Delta, \gamma}\, g_{\alpha \beta} + 
g_{\gamma \beta} \mathcal{D}^q _{r-\Delta,r} n^\gamma + 
g_{\alpha \gamma} \mathcal{D}^q _{r-\Delta,r} n^\gamma \right] e_{a}^\alpha\, e_b ^\beta,
\end{eqnarray}
where the projectors $e^{\alpha}_{a}$s are $e_t^\alpha\, \partial_\alpha=(\partial_t+\omega\,\partial_\phi)$,
 $e_\theta ^\alpha\,\partial_{\alpha}=\partial_\theta$ and $e_\phi^\alpha\,\partial_{\alpha}=\partial_\phi$.

Let us first determine $^{q} K^{+}_{tt}$, where $+$ denotes that the variable is 
associated with the external spacetime. Note that since $e_t^\alpha \partial_\alpha=(\partial_t+\omega \partial_\phi)$, one get the only
contribution from  $^qK^{+}_{tt}=(1/2)\,{ ^q \lie_n (g^{+}_{tt})}$. The other contribution to $^qK^{+}_{tt}$ from $g^{+}_{t\phi}$
is neglected since
$g^{+}_{t \phi}=-(2Ma\sin^2 \theta/r)$, is directly proportional to $a$ and further, 
together with $\omega=(2Ma/r^3)$ contributes an overall $a^2$ term. Note that due to 
the form of the normal vector, 
and the metric, only the first term in expansion in \eqref{qlie_expansion} contributes. Using 
the expression for $ \mathcal{D}^q _{r-\Delta,r}(r^{-1})$ in the appendix, eqn. \eqref{drminus1}, we get 
%
%
\begin{eqnarray*}
{}^{q}K^{+}_{tt}
&=&- \frac{M}{R_{0}^2}\left(1-\frac{2M}{R_{0}}\right)^{1/2}  \left[ 1+2 \frac{1-q}{2-q} \frac{\Delta}{R_0}+\dots\right],
\end{eqnarray*} 
and hence, using the metric, one easily determines that
\begin{eqnarray}
{}^q K_t^{+t} 
&=&\frac{M}{R_{0}^2}\left(1-\frac{2M}{R_{0}}\right)^{-1/2} \left[ 1+2 \frac{1-q}{2-q} \frac{\Delta}{R_0}+\dots\right].
\end{eqnarray}
%
%
%
Similarly, one finds the contribution from $^q K_{t\psi}$ as follows:
\begin{eqnarray}\label{ktpsi}
^q K^{+}_{t \psi}=(1/2)\, {^q \lie_n (g_{ab}^{+})} \, e_t^ a\, e_\psi^ b 
=(1/2)\, {^q \lie_n(g_{\phi t}^{+})} +(\omega/2)\, {^q \lie_n(g_{\phi \phi}^{+})}.
\end{eqnarray}
%
%
Using $\mathcal{D}^q _{r-\Delta ,r}(r^{-1})$ in the appendix, eqn. \eqref{drminus1}, we get that
\begin{eqnarray}\label{gtphi}
^q \lie_n (g_{\phi t}^{+})= n^r \mathcal{D}^q _{r-\Delta ,r} (g_{\phi t}^{+}) 
=\frac{2Ma \sin^2 \theta}{R_0^2} \left( 1+2 \frac{1-q}{2-q} \frac{\Delta}{R_0}+\dots \right).
\end{eqnarray}
Again, using $\mathcal{D}^q_{r-\Delta,r}(r^2)$ in the appendix , eqn. \eqref{dr2}, we get
\begin{eqnarray}\label{gphiphi}
^q \lie _n (g_{\phi \phi}^{+}) &=&n^r \mathcal{D}^q _{r-\Delta ,r}(r^2 \sin^2 \theta) \nonumber\\
&=&2R_0\sin^2 \theta \left(1-\frac{2M}{R_{0}}\right)^{1/2} \left(1-\frac{1-q}{2-q} \frac{\Delta} {R_0}+\dots\right).
\end{eqnarray}
%
Putting $\omega=(2Ma/R_{0}^3)$ in eqn. \eqref{ktpsi}, and using equations \eqref{gtphi} and \eqref{gphiphi}, we get 
\begin{eqnarray*}
^q K_{t \psi}^{+}=\frac{3Ma\sin^2 \theta }{R_0 ^2}\left(1-\frac{2M}{R_0}\right)^{1/2}\left[1+
\frac{1-q}{3-q} \left(\frac{\Delta}{R_0} \right)^2 +\dots \right].
\end{eqnarray*}
The expression naturally leads to the following expressions for extrinsic curvatures: 
\begin{eqnarray}
{^q K^{+t}_\psi}= {g^{tt}}\,\left( {^q K_{t \psi}^{+}}\right)= \frac{-3Ma \sin^2 \theta}{R_0^2}\left(1-\frac{2M}{R_0}\right)^{-1/2}\left[1
+ \frac{1-q}{3-q} \left(\frac{\Delta}{R_0} \right)^2 +\dots \right], \\
{^q K^{+\psi} _t}=g^{\psi \psi} \,\left({^q K_{t \psi}^{+}}\right)=
\frac{3Ma}{R_0 ^4}\left(1-\frac{2M}{R_0}\right)^{1/2} \left[1+ \frac{1-q}{3-q} \left(\frac{\Delta}{R_0} \right)^2 +\dots\right].
\end{eqnarray}
The angular components of the extrinsic curvatures are $^q K^{+}_{\theta \theta}$ and $^q K^{+}_{\psi \psi}$ and their expressions may be found
in exactly the same method and we get:
%
\begin{equation}
^q K_\theta ^{+\theta}= \, ^q K_\psi ^{+\psi}=\frac{1}{R_0} \left(1-\frac{2M}{R_0}\right)^{1/2}\left( 1-\frac{1-q}{2-q} \frac{\Delta}{R_0}
+\dots\right).
\end{equation}

For interior spacetime, we take it to be the flat Minkowski spacetime. So, to the past of the hypersurface at 
$r=R_{0}$, the spacetime $\mathcal{M^{-}}$
is given by the metric
\begin{equation}
ds^2=-\Big( 1-\frac{2M}{R_{0}}\Big)dt^2+d\rho^2+\rho ^2 d\Omega^2 
\end{equation} 
where $\rho$ is a radial coordinate. The intrinsic metric on the hypersurface from the interior matches
with the induced metric from the exterior region.
The normal to the hypersurface is $n^ \alpha= (\partial/\partial\rho)^{\alpha}$. 
The expressions for the extrinsic curvatures may be determined and the only non- vanishing components are $^q K_{\theta \theta}$ and $^q K_{\phi \phi}$:
\begin{eqnarray}
^q K_\theta ^{-\theta}=\,^q K_\psi ^{-\psi}=\frac{1}{R_0}  \Big( 1-\frac{1-q}{2-q} \frac{\Delta}{R_0}+............\Big)
\end{eqnarray} 

Let us now determine the stress- energy tensor of the thin shell of matter forming the hypersurface joining the two spacetimes.
The discontinuities in the extrinsic curvatures are related to the shell's surface stress-energy tensor $S^{ab}$. 
\begin{eqnarray}
8 \pi S^t_t = \Big[{^q K _\theta ^\theta }\Big] +\Big[{^q K_\psi ^\psi }\Big]\\
-8 \pi S^t_\psi=\Big[ {^q K _\psi ^t}\Big]\\
-8 \pi S^\psi _t=\Big[ {^q K ^\psi _t}\Big] \\
8 \pi S^\theta _\theta = \Big[{^q K _t ^t }\Big] +\Big[{^q K_\psi ^\psi }\Big]
\end{eqnarray}
The shell's matter may be assumed to be made of perfect fluid, with density  $\sigma=-S ^t _t$, pressure $p=S_\theta ^\theta$
and rotating with angular velocity $\omega=-S^t _\psi/(-S^t _t +S ^\psi _ \psi)$.
The expressions for these components of the energy momentum tensor are:
\begin{equation}
S_ \psi ^t=\frac{3Ma \sin^2 \theta}{8 \pi R_{0}^2}\left(1-\frac{2M}{r}\right)^{-1/2} \left[1
+ \frac{1-q}{3-q} \left(\frac{\Delta}{R_0} \right)^2 +\dots\right],
\end{equation}
\begin{equation}
S_t ^\psi =\frac{-3Ma}{8 \pi R_{0}^4} \left(1-\frac{2M}{r}\right)^{1/2} \left[1+ \frac{1-q}{3-q} \Big(\frac{\Delta}{R_0} \Big)^2 +\dots\right],
\end{equation}
%
%
%
%
%
%
%
%
%
%
\begin{eqnarray}
S_t ^t =\frac{-1}{4 \pi R_0}\Big(1-\sqrt{1-2M/R_0}\Big) \left[ 1-\frac{1-q}{2-q} \frac{\Delta}{R_0}+\dots\right],
\end{eqnarray}
%
%
%
%
%
%
%
%
\begin{eqnarray}
S_\theta ^\theta &=&\left[\frac{(1-2M/R_0)^{-1/2}}{8 \pi R_0}\right]\left[1-M/R_0-\sqrt{1-2M/R_0}\right]
-\left[\frac{(1-2M/R_0)^{-1/2}}{8 \pi R_0}\right]\nonumber\\
&&\left[1-4M/R_0+
\sqrt{1-2M/R_0}\right]\Big(\frac{1-q}{2-q} \Big) \Big(\frac{\Delta}{R_0} \Big)+\dots.\\
S_\theta ^\theta &=& S_\psi^\psi .
\end{eqnarray}
Quite naturally, all the expressions of the energy momentum tensor are modified due to the improved notion of fractional differential. The 
modification takes the thickness of the shell into account. One very interesting notion is the determination of the angular
velocity of the shell. The angular velocity is obtained from $\omega=S^t _\psi/(S^t _t -S ^\psi _ \psi)$. This gives for $R_0\gg 2M$,
\begin{equation}
\omega _{shell}=\frac{3a}{2R_0 ^2}+\frac{3a}{2MR_0}\frac{1-q}{2-q} \frac{\Delta}{R_0}+\dots.
\end{equation}
This expression given above for the angular velocity is different from that obtained in the usual case \cite{poisson} 
but reduces to it in the limit $\Delta/R_{0}\rightarrow 0.$

\subsection{Matching the Schwarzschild and the de-Sitter spacetimes}
The joining of exterior spacetime of the Schwarzschild black hole (taken as the exterior spacetime) with the de- Sitter spacetime
has been the subject of many investigations, which were particularly directed to create singularity free models
of black hole interior. One particularly interesting application was considered by 
Frolov, Markov and Mukhanov \cite{Frolov_markov}
to exemplify their \emph{limiting curvature hypothesis}. They suggested that inside the Schwarzschild black hole, very close to the
singularity, when the Planck scale is reached, there would be corrections to the Einstein theory of gravity. These corrections would 
not allow the curvature of the spacetime to dynamically grow to infinite values. Instead, the effective curvature of the 
spacetime would be bounded from below by $\ell_{p}^{-2}$, where $\ell_{p}$ is the Planck length. Naturally, this hypothesis
implies that there will be no curvature singularity. Instead, the model in \cite{Frolov_markov} proposes that very close to the spacetime singularity, where 
the curvature reaches the $\ell_{p}^{2}$,
the spacetime makes a transition from the Schwarzschild to the de- Sitter spacetime by passing through a very thin transition layer
\begin{figure}
\begin{center}
\includegraphics[width=7cm,height=6cm]{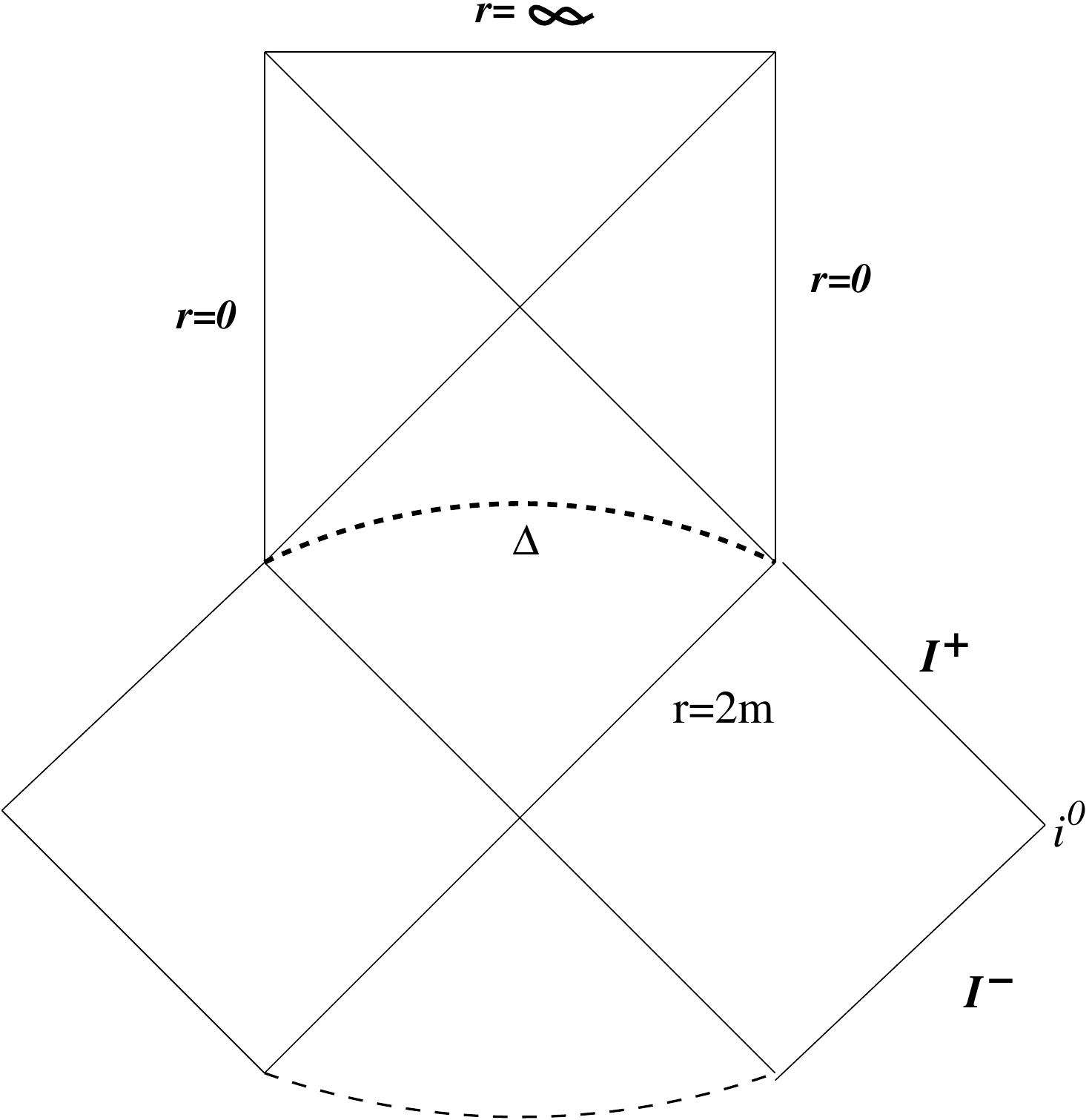}
\caption{The Frolov, Markov and Mukhanov model in which the Schwarzschild
black hole has a de Sitter world
in the interior. The spacelike hypersurface $\Delta$ represents the matching hypersurface
joining the two spacetimes. The $I^{+}$, $I^{-}$ and $i^{0}$ represent the future null, past null and the spatial infinities.}
\label{}
\end{center}
\end{figure}.
The spacetime passes through a deflation stage and instead of singularity, reaches a new inflating universe free of singularity. 
The matching of these two spacetimes require stress- energy tensors on the joining shell which violate energy conditions.

In the following, we recalculate the stress- energy tensor on the matching shell using the fractional calculus 
and show that the stress- tensor is modified. The modified stress tensor will be shown to lead to smooth matching
of the spacetimes.  Let us match the de-Sitter spacetime with the interior Schwarzschild spacetime. 
The metric for the two spacetimes may be 
written in a combined form as: 
\begin{eqnarray}
ds^2 =f(r)\,dv^2 + 2dv dr +r^2\,d \Omega^2,
\end{eqnarray}
where $f(r)=(2M/r -1)$ for the Schwarzschild metric and the 
$f(r)=\left[(r/l)^2 -1 \right]$ for the de-Sitter metric.
For simplification, let us define a new set of coordinates: $v=\lambda/\sqrt{f}$. The induced metric on 
the spacelike surface becomes $ds^2=d\lambda^2 + r^2 d \Omega^2$. The normal to this surface is given by:
\begin{equation}
n _\alpha =\left[0, - \frac{1}{\sqrt{f}},0,0\right],
\end{equation}
and $n^\alpha = \left(1/\sqrt{f},\sqrt{f},0,0\right)$. The coordinates in the spacetime is taken to be $x^{\alpha}=(v,r, \theta,\phi)$ 
and that of the hypersurface to be $y^{a}=(\lambda,\theta,\phi)$. This implies that 
$e^{\alpha}_{q}\partial_{\alpha}=(1/\sqrt{f})(\partial/\partial v)^{\alpha}$,
$e^{\alpha}_{\theta}\partial_{\alpha}=(\partial/\partial \theta)^{\alpha}$ 
and $e^{\alpha}_{\phi}\partial_{\alpha}=(\partial/\partial \phi)^{\alpha}$.

Let us evaluate the extrinsic curvatures.
%
%
The general expressions for these quantities for either of these spacetimes are given by the following:
 \begin{eqnarray}
 ^q K_{qq} &=&\frac{1}{2\sqrt{f}} \,\mathcal{D} ^q _{r-\Delta,r} (g_{vv}), \\
 ^q K_{\theta \theta} &=&(\sqrt{f}/2) \, \mathcal{D} ^q _{r-\Delta,r} (g_{\theta \theta}),\\
 ^q K_{\phi \phi} &=&(\sqrt{f}/2) \, \mathcal{D} ^q _{r-\Delta,r} (g_{\phi \phi}).
 \end{eqnarray}

For the Schwarschild Metric, which is taken to be the interior spacetime, these expressions are
obtained using the equations \eqref{drminus1} and \eqref{dr2} are:
\begin{eqnarray}
   ^q K_q ^{q-}&=&- \frac{M}{R_0 ^2} \left(\frac{2M}{R_0^2}-1 \right)^{-1/2} \Big[ 1+2 \frac{1-q}{2-q} \frac{\Delta}{R_0}+\cdots\Big],\\
   ^q K_{\theta}^{\theta-}&=& {}^q K_{\phi}^{\phi-} =\frac{1}{R_0}  \left(\frac{2M}{R_0}-1\right)^{1/2} \Big[1-\frac{1-q}{2-q} \frac{\Delta}{R_0}+\cdots \Big].
\end{eqnarray}
%
%

For the de-Sitter metric, taken to be the external or the future spacetime, the same expressions are given as,
using \eqref{dr2} are :
\begin{eqnarray}
 ^q K_{q}^{q+}& =&\frac{R_0}{l^2}\left[{\Big(\frac{R_{0}}{l}\Big)^2-1}\right]^{-1/2}\Big[1-\frac{1-q}{2-q} \frac{\Delta}{R_{0}}
 +\cdots\Big],\\
^q K_{\theta}^{\theta+}& =& ={}^q K_{\phi}^{\phi+}=\frac{1}{R_{0}} \left[\Big( \frac{r}{l}\Big)^2-1\right]^{1/2} \Big[1-\frac{1-q}{2-q} \frac{\Delta}{R_{0}}+\cdots\Big].
\end{eqnarray}
%


The jump in the components of the extrinsic curvatures are given by:
 \begin{eqnarray}
 \kappa =[^q K_q ^q ]&=&\frac{R_0}{l^2}\left[{\Big(\frac{R_{0}}{l}\Big)^2-1}\right]^{-1/2}\Big[1-\frac{1-q}{2-q} \frac{\Delta}{R_0}+\cdots \Big] 
 \nonumber\\
 &&~~~~~~~~~~~~~~~+ \frac{M}{R_0 ^2} \Big(\frac{2M}{R_0^2}-1 \Big)^{-1/2} \Big[ 1+2 \frac{1-q}{2-q} \frac{\Delta}{R_0}+\cdots\Big],\\
 \lambda =[^q K_\theta ^\theta]&=&-\frac{1}{R_{0}^{2}}\left(\frac{2M}{R_0}-1\right)^{1/2} \Big[1-\frac{1-q}{2-q} \frac{\Delta}{R_0}+\cdots
 \Big]\nonumber\\
 &&~~~~~~~~~~~~~~~+\frac{1}{R_{0}}\left[\left(\frac{R_{0}}{l}\right)^2-1\right]^{1/2}  \left[1-\frac{1-q}{2-q} \frac{\Delta}{R_0}+\cdots\right].
 \end{eqnarray}
 The components of the stress- energy tensor is given by $ S_q ^q =\lambda/4 \pi$ and 
$ S_\theta^\theta =S_\phi^ \phi=(\kappa +\lambda)/8 \pi$. Quite noticably, the values of the energy momentum tensors
are markedly different from those obtained in \cite{Frolov_markov}. The values differ by quantities which 
are proportional to the ratio $(\Delta/R_{0})$, and hence by choosing the value of this ratio judiciously,
it can be easily seen that the energy momentum tensor can be made to vanish. Hence, one may match 
the two spacetimes smoothly across a spacelike hypersurface.


\section{Junction conditions for null hypersurfaces}
Let us consider a null hypersurface that partitions the $4$- dimensional spacetime into two regions $\left[\mathcal{M}^{+}, g_{\mu\nu}^{+}(x^{+})\right]$ and $\left[\mathcal{M}^{-}, g_{\mu\nu}^{-}(x^{-})\right]$, which we shall conveniently call as the future and the past respectively. Let us denote the coordinates of the spacetime as $x^{\alpha},\, \alpha=0,1,2,3$, whereas the coordinates on either side of the hypersurface will be denoted by $y^{a},\,a=1,2,3$, which will mean the collective coordinates $(\lambda,\theta^{A})$, where $\theta^{A}, A=(2,3)$ denotes
the variables on the two- dimensional cross- sections of the hypersurface. 
On each side of the hypersurface, one may construct the tangents to the generators of the null hypersurface ($\ell^{\alpha}$) and the transverse
spacelike vectors ($e^{\alpha}_{A}$), which are tangents to the cross-sections
(taken to be compact) of the hypersurface. These vectors shall be denoted by:
\begin{equation}
l^{\alpha}=e^{\alpha}_{\lambda}=\left(\frac{\partial x^{\alpha}}{\partial \lambda}\right)_{\theta^{A}};  \hspace*{0.4cm} e^{\alpha}_{A}=\left(\frac{\partial x^{\alpha}}{\partial \theta^{A}}\right)_{\lambda},
\end{equation}
with the following properties: $
\ell^{\alpha}\ell_{\alpha}=0, \hspace{1pt} \ell_{\alpha}e^{\alpha}_{A}=0$.
These vectors may be constructed for both sides of the null hypersurface. Further, on each side, the basis needs four vectors and the fourth vector, will be taken to be a null vector. It will be denoted by $n^{\alpha}$ with the following properties: $\ell^{\alpha}n_{\alpha}=-1,\hspace{1pt} n^{\alpha}n_{\alpha}=0, \hspace{1pt} n_{\alpha}e^{\alpha}_{A}=0$. 

The typical situation with a null surface is that the usual extrinsic
curvature, $K_{ab}=(1/2)\left(\pounds_{\ell}\,g_{\alpha\beta}\right)
e^{\alpha}_{a}\,e^{\beta}_{b}$, corresponding to the normal to the hypersurface is continuous, since the normal
is also the tangent $\ell^{\alpha}$. So, one usually defines
the transverse component of the extrinsic curvature 
corresponding to the null vector field normal to the transverse cross- sections of the hypersurface. This vector
is $n^{a}$, such that $\ell.n=-1$. The transverse extrinsic curvature 
may be defined as $C_{ab}=(1/2)\left(\pounds_{n}\,g_{\alpha\beta}\right)
e^{\alpha}_{a}\,e^{\beta}_{b}$. The stress- energy tensor of the shell is given by:
$S^{\alpha\beta}=\mu\,\ell^{\alpha}\ell^{\beta}+p\,\sigma^{AB}\,e^{\alpha}_{A}\,
e^{\beta}_{B}$,
where $\mu=(-1/8\pi)\sigma^{AB}[C_{AB}]$ is the shell's surface
density  and $p=(-1/8\pi)[C_{\lambda\lambda}]$ is the surface pressure.

%
%
%
%
%
%
%


\subsection{Null Charged Shell Collapsing on a Charged Black Hole}

Let us consider a spherically symmetric charged black hole of mass $M$ and charge $Q$ on
which a null charged shell of mass $E$ and charge $q$ collapses. Outside
the shell, the spacetime outside the total configuration may be viewed as
a spherically symmetric Reissner- Nordstrom type geometry with mass $(M+E)$
and charge $(Q+q)$. As before, the spacetimes outside and inside the shell shall
be denoted by $+$ and $-$ respectively.
\begin{eqnarray}
ds^2=-f_{\pm} (r)+dv^2 +2dv dr +r^2 d\Omega^2 \\
f_+(r)=1-\frac{2(M+E)}{r}+\frac{(Q+q)^2}{r^2} \\
f_-(r)=1-\frac{2M}{r}+\frac{Q^2}{r^2}.
\end{eqnarray}
The coordinates of the spacetime is given by $x^{\alpha}=(v,r,\theta,\phi)$. 
The surface of the shell is given by $v=v_{0}$ with coordinates of the shell
being $y^{a}=(r,\theta,\phi)$. The vector fields are given by $e^{\alpha}_{r}\partial_{\alpha}\equiv\ell^{\alpha}=
-(\partial/\partial r)^{\alpha}$, $e^{\alpha}_{\theta}\partial_{\alpha}=(\partial/\partial \theta)^{\alpha}$,
$e^{\alpha}_{\phi}\partial_{\alpha}=(\partial/\partial \phi)^{\alpha}$.
Note that the vector $\ell^{\alpha}$ is the generator of the null surface.
The transverse null vector required to complete the basis is $n{^\alpha}= \left\{f_{\pm}(r)/2\right\}(\partial/\partial r)^{\alpha}.$

The metric is continuous across the shell. Let us check that the extrinsic curvature of the null surface corresponding to the
 null normal $\ell^{\alpha}$ is also continuous on either side of the surface. This is always true for the non- fractional case
 and precisely for this reason, the concept of the transverse curvatures have been introduced.
We show that for the fractional case too, the extrinsic curvatures corresponding to the null normal
of the surface is continuous.
For the interior solution, we get that $\ell^{-\alpha}=(0,1,0,0)$ and hence, the components of the fractional extrinsic curvature
 on the cross- sections are: $^q K_{AB}^- =
(1/2){{}^q\lie_\ell\, g_{AB}}$.
%
Using the formulae from the appendix, equation \eqref{dr2}, we get 
\begin{eqnarray*}
^q K_{\theta \theta}^-= R_0 \Big[1-\frac{1-q}{2-q} \frac{\Delta}{R_0}+\cdots\Big]\\
^q K_{\phi \phi}^-=R_0 \sin^2 \theta \Big[1-\frac{1-q}{2-q} \frac{\Delta}{R_0}+\cdots \Big]
\end{eqnarray*}
These two equations may be combined to the following form:
\begin{equation}
^q K_{AB}^-= {R_0 ^2} \Big[1-\frac{1-q}{2-q} \frac{\Delta}{R_0}+\cdots \Big]\sigma_{AB}
\end{equation}
For the exterior solution too, the null normal is given by $\ell^{+\alpha}=(0,1,0,0)$ and the extrinsic curvature
corresponding to this null normal is $^q K_{AB}^+=\frac{1}{2} {^q \lie_\ell(g_{AB})}$ is given by:
\begin{equation}
^q K_{AB}^+= {R_0 ^2} \Big[1-\frac{1-q}{2-q} \frac{\Delta}{R_0}+\cdots \Big] \sigma_{AB}
\end{equation}
This implies that the extrinsic curvatures are also continuous $^q K_{AB}^+=^q K_{AB}^-$.

%

The transverse extrinsic curvature is not continuous for this metric. 
The expression for $^q C_{\theta \theta}^+$ is given by 
\begin{eqnarray}
^q C_{\theta \theta}^+&=&(1/2)\left[n^r \mathcal{D}^q_{r-\Delta,r} (g_{\theta \theta})\right]\nonumber\\
&=&f_{+}(r)\,R_{0}\Big[1-\frac{1-q}{2-q} \frac{\Delta}{R_0}+\cdots \Big] 
\end{eqnarray}
Similarly  for $^q C_{\phi \phi}^+$, we get:
\begin{equation}
^q C_{\phi \phi}^+= f_{+}(r) R_{0} \sin^2 \theta \left[1-\frac{1-q}{2-q} \frac{\Delta}{R_0}+\cdots \right] .
\end{equation}
So these two expressions may be combined to give:
\begin{equation}
^q C_{AB}^+=\frac{f_{+}(r)}{R_0} \left[1-\frac{1-q}{2-q} \frac{\Delta}{R_0}+\cdots \right] \sigma_{AB},
\end{equation}
where $\sigma_{AB}=R_0^2+R_0^2 \sin^2 \theta$.
Similarly, for the interior spacetime, the transverse component of the extrinsic curvature is given by:
\begin{equation}
^q C_{AB}^-=\frac{f_{-}(r)}{R_0} \left[1-\frac{1-q}{2-q} \frac{\Delta}{R_0}+\cdots \right] \sigma_{AB}  
\end{equation}
These equations immediately imply that the shell's surface pressure is zero for this case. The shell's surface density is 
\begin{eqnarray}
\mu =-\frac{1}{4 \pi R} \Big( \frac{2Qq+q^2}{R^2}-\frac{2E}{R} \Big) \left[1-\frac{1-q}{2-q} \frac{\Delta}{R}+\cdots \right].
\end{eqnarray}
This relation clearly implies that to satisfy the weak energy condition, we must have
 \begin{equation}
 2E \geq \frac{2Qq+q^2}{M+\sqrt{M^2 -Q^2}}.
 \end{equation}
As a simple application, let us study if the charged black hole may be overcharged, so that the total charge $(Q+q)$ exceed the total mass
$(M+E)$. It is a simple matter to check that the condition for overcharging violates the weak energy
condition. So, even in the fractional modification, a charged black hole cannot be overcharged.


\subsection{Matching Schwarschild and de-Sitter spacetimes across horizons}
Let's start with a general form of the metric and then we shall specialize to the individual cases. The general form for a spherical symmetric metric
in the advanced Eddington -Finkelstein coordinates is given by:
\begin{equation}
ds^2=-f(r)\,dv^2+ 2dv\,dr+r^2\, d \Omega^2.
\end{equation}  
The coordinates of the spacetime is given by $x^{\alpha}=(v,r,\theta,\phi)$. 
Let us assume a null hypersurface (a shell) given by $r=r_{0}$ with coordinates of the shell
being $y^{a}=(v,\theta,\phi)$. The null surface is foliated by compact surface $S^{2}$. The vector fields tangent to the sphere are given by, $e^{\alpha}_{\theta}=(\partial/\partial \theta)^{\alpha}$ $e^{\alpha}_{\phi}=(\partial/\partial \phi)^{\alpha}$. 
Let us now determine the set of null vectors tangent to the null surface which is 
given by the relation $f(r_{0})=0$. The generator of the null surface is $\ell^ \alpha = (\partial/\partial v)^{\alpha}$
and the transverse null vector is $n^ \alpha =-(\partial/\partial r)^{\alpha}$.

Let us consider the interior metric ($\mathcal{M}^{-}$) to be the de- Sitter spacetime:
\begin{equation}
ds^2_- =-\left[1- \left(\frac{r}{l} \right)^2 \right] dv^2+2 dv dr+r^2 d\Omega^2.
\end{equation}
As usual, the standard extrinsic curvatures associated to the null normals of the are continuous and hence let us calculate the
transverse extrinsic curvatures $ ^q C^- _{\theta \theta}$ and $^q C^- _{\phi \phi}$:
\begin{equation*}
^q C_{\theta \theta}^-=R_0  \left[1-\frac{1-q}{2-q} \frac{\Delta}{R_0}+\cdots \right],
\end{equation*}
and similarly  for $^q C_{\phi \phi}^+$ , we get
\begin{eqnarray*}
^q C_{\phi \phi}^-=R_0 \sin^2\theta \left[1-\frac{1-q}{2-q} \frac{\Delta}{R_0}+\cdots \right].
\end{eqnarray*}
So combining them together, we get:
\begin{equation}
^q C_{AB}^-=\frac{1}{R_0} \left[1-\frac{1-q}{2-q} \frac{\Delta}{R_0}+\cdots \right]\sigma_{AB},
\end{equation}
where $\sigma_{AB}=R_0^2+R_0^2 \sin^2 \theta$. The quantity $^q C^- _{vv}$ gives:
\begin{equation}
^q C_{vv}^-=\frac{1}{ a^2}R_0 \left[1-\frac{1-q}{2-q} \frac{\Delta}{R_0}+\cdots \right].
\end{equation}

The exterior spacetime is taken to be the Schwarzschild spacetime $\mathcal(M^{+})$, with the metric
\begin{equation}
ds^2_- =-\Big(1- 2m/r  \Big) dv^2+2 dv dr+r^2 d\Omega^2
\end{equation}
Again, the co-ordinate on null shell are $(v, \theta , \phi)$. Let us calculate transverse curvatures.
Just as in the previous case, the result is 
\begin{equation}
^q C_{AB}^+=\frac{1}{R_0} \left[1-\frac{1-q}{2-q} \frac{\Delta}{R_0}+\cdots \right]\sigma_{AB}.
\end{equation}
The $^q C^+_{vv}$ is given by:
\begin{equation}
^q C_{vv}^+=-m\frac{1}{R^2_0} \left[ 1+2 \frac{1-q}{2-q} \frac{\Delta}{R_0}+\cdots \right].
\end{equation}
Let us calculate the quantities associated with the shell. The surface density $\mu=0$. The pressure is  given by:
\begin{equation}
p=\frac{-1}{8 \pi}\left[\Big( \frac{R_0}{l^2}+\frac{m}{R_0 ^2}\Big) - \Big( \frac{R_0}{l^2} - \frac{2m}{R_0^2}\Big)\frac{1-q}{2-q}\frac{\Delta}{R_0} +\cdots \right].
\end{equation}
So, if the matching surface is the horizon, $R_{0}=2m=l$ and hence the pressure must by non vanishing. However, in the fractional modification,
we may choose the $\Delta/R_{0}$ judiciously to get a smooth matching of the two spacetimes.
 
\subsection{Matching the Reissner- Nordstrom and the de-Sitter spacetimes}
Let us determine the criteria for matching the Reissner- Nordstrom spacetime and the de- Sitter spacetimes on the inner horizon of the non- extremal charged black hole. Interestingly the matching is to be carried out on the inner horizon as
was firts proposed in \cite{Barrabes_Israel}. The Penrose diagram is given in fig 2.

\begin{figure}
\begin{center}
\includegraphics[width=4.5cm,height=5.5cm]{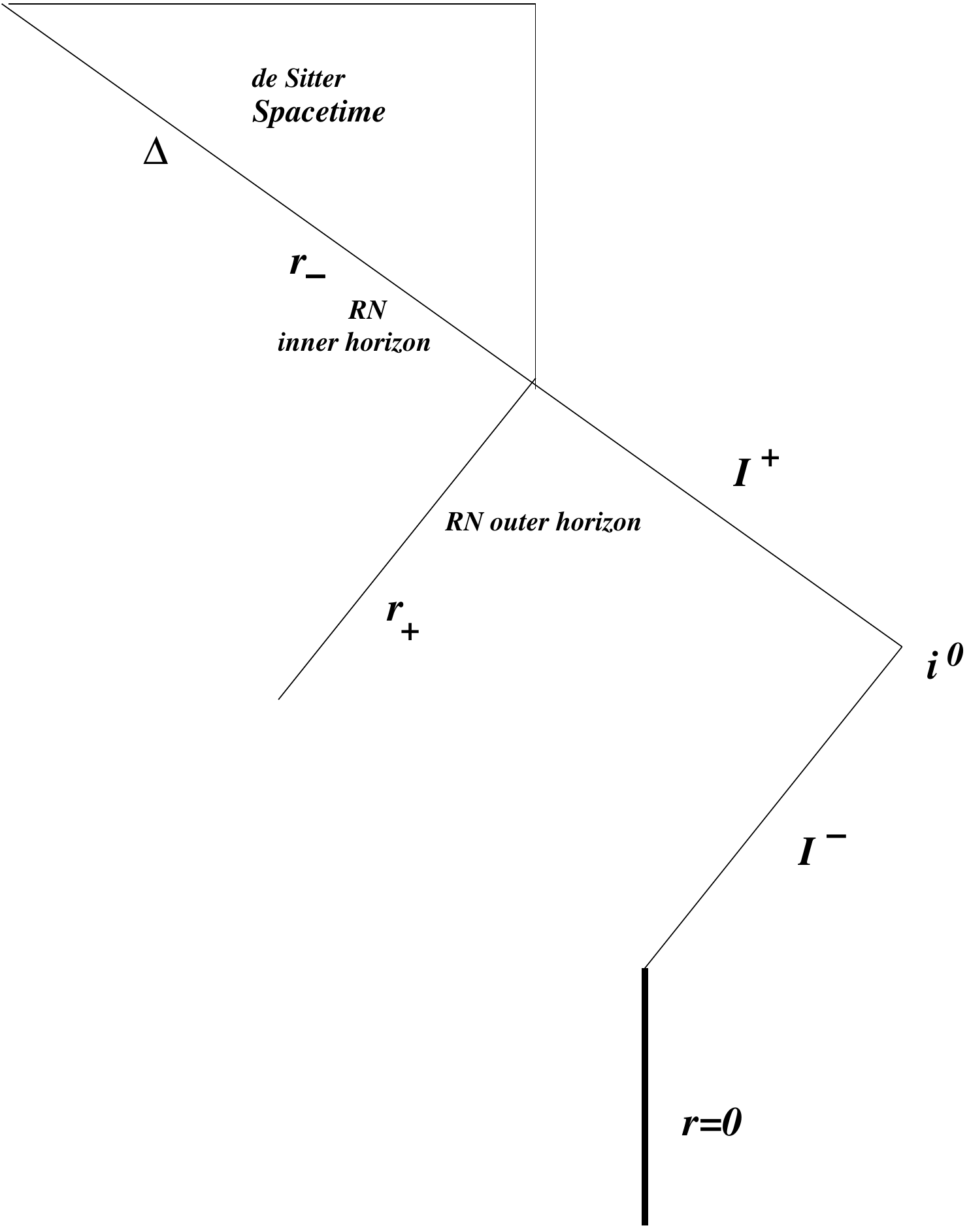}
\caption{The Barrabes- Israel model in which the inner horizon of the
non -extremal Reissner- Nordstrom 
black hole is joined to a de Sitter world
in the interior. The null surface $\Delta$ represents the matching hypersurface
joining the two spacetimes. The $I^{+}$, $I^{-}$ and $i^{0}$ represent the future null, past null and the spatial infinities.}
\label{}
\end{center}
\end{figure}.

The external spacetime is the Reissner- Nordstrom spacetime ($M^{+}$) with the following metric: 
\begin{equation}
ds^2_- =-\Big(1-\frac{2m}{r} + \frac{Q^2}{r^2} \Big) dv^2+2 dv dr+r^2 d\Omega^2.
\end{equation}
The co-ordinates on null shell are $(v, \theta , \phi)$ and $R_0=m-\sqrt{m^2-Q^2}$.
The transverse extrinsic curvatures $ ^q C^+ _{\theta \theta}$ and $^q C^+_{\phi \phi}$ may be written as
\begin{equation}
^q C_{AB}^+=\frac{1}{R_0} \left[1-\frac{1-q}{2-q} \frac{\Delta}{R_0}+\cdots \right]\sigma_{AB}.
\end{equation}
\begin{equation}
^q C_{vv}^+=\left[ \left(\frac{Q^2}{R_0^3}- \frac{m}{R_0^2}\right)- 
\frac{1-q}{2-q} \frac{\Delta}{R_0} \left(\frac{2m}{R_0^2}-\frac{3 Q^2}{R^3} \right)+\cdots \right].
\end{equation}
The interior spacetime is the de- Sitter spacetime with matching at $R_{0}=l$. 
The curvatures have already been found out in the previous subsection.
The properties of the shell may be immediately obtained. The surface density $\mu=0$ but the pressure is
\begin{equation}
p=\frac{-1}{8 \pi}\left[  \left( \frac{R_0}{l^2}+\frac{m}{R_0 ^2}-\frac{Q^2}{R_0 ^3}\right) - \left( \frac{R_0}{l^2} - \frac{2m}{R_0^2} +\frac{3Q^2}{R_0^2}\right)\frac{1-q}{2-q}\frac{\Delta}{R_0} +\cdots \right].
\end{equation}
So, again, in the standard case, when $(\Delta/R_{0})=0$ the spacetimes matching requires a shell which shall hold this pressure and hence 
the matching is not smooth. Incidentally, in \cite{Barrabes_Israel}, the
authors noted that for special case like $3l^{2}=Q^{2}$, there is a smooth matching of the two spacetimes. This matching
is a special case. The fractional generalisation however, shows that it is possible to adjust the parameter $(\Delta/R_{0})$
to get a vanishing pressure and hence, a smooth matching of the spacetimes on the hypersurface.

\section{Discussions}
In this paper, we have developed the fractional generalisation of the Israel- Darmois- Lanczos
junction conditions for spacelike/timelike as well as for null hypersurfaces. We have observed that
there is a significant modifications due to the fractional generalisation. First, due to the definition
of the fractional differentiation through an integral, it automatically incorporates the non- local spacetime correlations
into itself. As a manifestation of this, the thickness of the shell gets incorporated into to the 
values of the shell's properties like the energy and pressure. We have taken several examples
and have demonstrated that by choosing this thickness parameter $\Delta/R_{0}$ judiciously,
it is possible to join many spactimes smoothly across spacelike timelike or null hypersurfaces.

A point of crucial importance is that must be mentioned here is that the dimension of the spacetime 
has been taken to be integral. Fractal dimensions may also be possibly included. 
In fact, general relativity may also be suitably adapted for 
 \emph{fractal spacetimes}, which would also require revising our notions of coordinate transformations and covariance. However, we 
 have not attempted this path, of altering the theory of general relativity to recast it for all spacetime dimensions, integral or 
 non- integral. Instead, we have looked for alternate avenues by generalising the notion of Lie- derivative which is 
initrinsically attached to the differentiable structure of the spacetime \cite{Krisch}. Unlike the usual Levi-Civita connection, 
there is no requirement of the metric and hence the Lie derivative is much primitive and turns out to be most useful. 
This generalisation has been used to construct the extrinsic curvatures and hence the surface properties of the shell.
In the appendix, we have developed the reasons as to why we should expect that there should be some modification
in the dynamics as well. We show that the Einstein equations modify significantly. The ramifications of these 
issues shall be dealt with in future papers.

\section*{Acknowledgements}
The authors acknowledge fruitful discussions with Amit Ghosh.

 \section{Appendix}
 \subsection{Fractional derivative}
 The Riemann- Liouville definition of fractional calculus is usually given in the form of a integral transform of a specialised type, as given below
 \cite{oldham_spanier, miller_ross,samco_kilbas_marchev}:
\begin{equation}
D_{x}^{-\nu}f(x)=\frac{1}{\Gamma(\nu)}\int_{a}^{x}\, (x-y)^{\nu-1}f(y)\,dy,
\end{equation}
where $\nu > 0$. This definition is the foundation of the theory of fractional differentiation (and integration), but breaks down at the integral points, $\nu=0,-1,-2, \cdots$. At those points the integration may however be replaced by the ordinary integration formula.     

The Caputo derivative is a modification of the Riemann- Liouville derivative 
where suitable modification have been applied so that it satisfies all the rules of a derivative. The Caputo derivative is defined as follows
 \cite{oldham_spanier, miller_ross,samco_kilbas_marchev}:
\begin{equation}
D_{x}^{q}f(x)=\frac{1}{\Gamma(1-q)}\int_{a}^{x}\, (x-y)^{-q}\, \frac{\partial f(y)}{\partial y}\,dy,
\end{equation}
where the superscript $q$ denotes the fractional parameter, $0<q\le 1$.
To take into account of the tensor indices, a further modification is added in \cite{Krisch} as follows:

 \begin{equation}
\mathcal{D}^q_{x,k}(x^{\prime\, i})=\frac{\Gamma{(2-q)}}{\Gamma{(1-q)}(\Delta)^{1-q}} \int_{x}^{x^{\prime}} \frac{\partial y^{i}}{\partial y^{k}} (x^{\prime}-y)^{-q}dy.
 \end{equation}
For example, if the integration of the metric variable $g_{ij}(r)$ is to be carried out from one end of the 
shell (of thickness $\Delta$) to the other, the above definition gives:
 \begin{equation}\label{fracderiv_def1}
\mathcal{D}^q_{r-\Delta,r}(g_{ij})=\frac{\Gamma{(2-q)}}{\Gamma{(1-q)}(\Delta)^{1-q}} \int_{r-\Delta}^{r} \frac{\partial g_{ij}(w)}{\partial w} (r-w)^{-q}dw,
 \end{equation}
where the integration limits have been chosen appropriately. This definition has been utilised in this paper.

\subsection{Beta Function and relation to the Hypergeometric functions}
In the paper, we have frequently made use of the Beta function, defined as:
\begin{equation}
B_x(a,b)=\int_{0}^{1}t^{a-1}(1-t)^{b-1} dt,
\end{equation}
where $a>0,b>0$. In general, we may also write it as 
\begin{equation}
B_x(a,b)=x^a\left(\frac{1}{a}+\frac{1-b}{1+a}x+\cdots\right),
\end{equation}
and hence the above equation implies naturally that:
\begin{equation}\label{1minusq2}
B_{\frac{\Delta}{R_0}}(1-q,2)=\left(\frac{\Delta}{R_0} \right)^{1-q} \frac{1}{1-q} \left [1-\frac{1-q}{2-q} \frac{\Delta}{R_0}+\cdots\right].
\end{equation}
If, $(\Delta/R) \neq 1$, we may also use the relation between Beta function and Hypergeometric function:
\begin{equation}
B_x(a,b)=\left(x^a/a\right) {{}_2F_{1}}(a,b,c;x)=\left(x^a/a\right) {_2F_{1}}(a,b-1,a+1;x),
\end{equation}
where 
\begin{equation}
{_2F_{1}}(a,b,c;x)=1+\frac{ab}{c}x+\frac{a(a+1)b(b+1)}{c(c+1)}\frac{x^2}{1}+\cdots.
\end{equation}
This gives the following two useful forms:
\begin{equation}\label{1minusqminus1}
B_{\frac{\Delta}{R_0}}(1-q,-1)=\frac{(\Delta/R_0)^{1-q}}{1-q} \Big[1+2 \frac{1-q}{2-q} \frac{\Delta}{R_0}+6 \frac{1-q}{3-q} \Big(\frac{\Delta}{R_0} \Big)^2+\cdots\Big]
\end{equation}
 \begin{equation}\label{1minusqminus2}
 B_{\frac{\Delta}{R_0}}(1-q,-2)=\frac{(\Delta/R_0)^{1-q}}{1-q} \Big[1+3 \frac{1-q}{2-q} \frac{\Delta}{R_0}+12 \frac{1-q}{3-q} \Big(\frac{\Delta}{R_0} \Big)^2+\cdots\Big].
 \end{equation}
These equations have been used below to derive the results used in the main text. Let us first evaluate $\mathcal{D}^q_{r-\Delta,r}(r^2)$.
Using eqn. \eqref{fracderiv_def1}, we get:
 \begin{equation}
 \mathcal{D}^q_{r-\Delta,r}(r^2)=\frac{2\,\Gamma{(2-q)}}{\Gamma{(1-q)}(\Delta)^{1-q}} \int_{r-\Delta}^{r} w (r-w)^{-q}dw, 
 \end{equation}
 where the integration limit is chosen to take the thickness of the hypersurface into account. Using
 the change of variables, $r-w=t$, the limit also changes from $r-\Delta$ to $\Delta$ and $r$ to $0$.
 This gives us:
  \begin{equation}\label{dr2}
 \mathcal{D}^q_{r-\Delta,r}(r^2)= \frac{2\Gamma{(2-q)}}{\Gamma{(1-q)}\Delta^{1-q}} \int_{0}^{\Delta} (r-t) t^{-q}dt.
 \end{equation}
Again, make a change of variables $t/r=y $ and also put $r=R_0$ as we match on the hypersurface placed at $r=R_0 $. 
\begin{eqnarray}
 \mathcal{D}^q_{r-\Delta,r}(r^2)&=& \frac{2\,\Gamma{(2-q)}}{\Gamma{(1-q)}\Delta^{1-q}} R_0^{2-q} \int_{0}^{\frac{\Delta}{R_0}} 
  (1-y) y^{-q}dy\\
  &=& \frac{2\,\Gamma{(2-q)}}{\Gamma{(1-q)} \Delta^{1-q}} R_0^{2-q} B_{\frac{\Delta}{R_0}}(1-q,2). 
 \end{eqnarray}
 Using the form of eqn. \eqref{1minusq2} and  property of Gamma function $(1-q) \Gamma (1-q)=\Gamma (2-q)$, we get,
 \begin{equation}
 \mathcal{D}^q_{r-\Delta ,r}(r^2)=2R_0 \left[1-\frac{1-q}{2-q} \frac{\Delta}{R_0}+\cdots\right].
 \end{equation}
For $\mathcal{D}^q_{r-\Delta,r}(r^{-1})$, a similar calculation yields the following result:
 \begin{eqnarray}
 \mathcal{D}^q_{r-\Delta,r}(r^{-1})&=& \frac{\Gamma{(2-q)}}{\Gamma{(1-q)}\Delta^{1-q}} R_0^{-1-q} 
 \int_{0}^{\frac{\Delta}{R_0}}  (1-y)^{-2} y^{-q}dy,\\
 &=&- \frac{\Gamma{(2-q)}}{\Gamma{(1-q)}\Delta^{1-q}} R_0^{-1-q}\, B_{\frac{\Delta}{R_0}}(1-q,-1).
 \end{eqnarray}
 Using eqn. \eqref{1minusqminus1} and property of Gamma function i.e $(1-q) \Gamma (1-q)=\Gamma (2-q)$ we get:
 \begin{equation}\label{drminus1}
 \mathcal{D}^q_{r-\Delta ,r}(r^{-1})=-\frac{1}{R^2_0} \Big[ 1+2 \frac{1-q}{2-q} \frac{\Delta}{R_0}+\cdots \Big].
 \end{equation}
 The computation for $\mathcal{D}^q_{r-\Delta,r}(r^{-2})$ proceeds along similar lines and gives:
  \begin{eqnarray}
  \mathcal{D}^q_{r-\Delta,r}(r^{-2})&=& \frac{-2\,\Gamma{(2-q)}}{\Gamma{(1-q)}\Delta^{1-q}} R_0^{-2-q} \int_{0}^{\frac{\Delta}{R_0}}  (1-y)^{-3} y^{-q}dy\\
  &=& \frac{-2\,\Gamma{(2-q)}}{\Gamma{(1-q)}\Delta^{1-q}} R_0^{-2-q}\, B_{\frac{\Delta}{R_0}}(1-q,-2),
  \end{eqnarray}
  which using the equation \eqref{1minusqminus2} and $(1-q) \Gamma (1-q)=\Gamma (2-q)$ gives us:
 \begin{equation}\label{drminus2}
 \mathcal{D}^q_{r-\Delta ,r}(r^{-2})=-\frac{2}{R^3_0} \Big[ 1+3 \frac{1-q}{2-q} \frac{\Delta}{R_0}+\cdots \Big].
 \end{equation}
 \subsection{Modification of the Einstein equations}\label{appendix3}
 The fractional derivative leads to a modification of the partial derivative. 
 From the previous sections, we note that the Caputo derivative 
 modifies the derivative through a factor $ (1-q)\Delta/R_0$. Let
 us use this form to write for any function $g$,
 a modification of the derivative operator as:
\begin{equation} \label{frac}
\mathcal{D} g=\partial \mathbf{g} \left[ 1 \pm \beta (1-q) \frac{\Delta}{R_0} \pm \cdots \right]
\end{equation}
here $\beta$ is some constant, and $q$ denotes the fractional parameter. Using
this definition of the derivative, the relation between new Christoffel symbol (for
non- Levi-Civita connection) and old Christoffel symbol (for
Levi-Civita connection) becomes :
\begin{equation} \label{chris}
\tilde{\Gamma} ^{\alpha}_{ \beta \gamma}= \Gamma ^{\alpha}_{ \beta \gamma} \pm \tilde{\alpha}(1-q) \frac{\Delta}{R_0} + \cdots
\end{equation} 
where $\tilde{\alpha}$ is some constant. This gives a relation between old Riemann tensor and new Riemann tensor. The usual 
definition 
\begin{equation}
R^ \alpha _{\beta \gamma \delta}= \partial _\gamma \Gamma^\alpha _{\beta \delta}-\partial _\delta \Gamma ^\alpha _{\beta \gamma}+ \Gamma ^\nu_{\beta \delta} \Gamma ^\alpha _{\nu \gamma} -\Gamma ^\nu _{\beta \gamma} \Gamma ^\alpha _{\nu \delta}
\end{equation}
is modified to a new definition 
\begin{equation}
\tilde{R}^ \alpha _{\beta \gamma \delta}= \mathcal{D} _\gamma \tilde{\Gamma} ^\alpha _{\beta \delta}-\mathcal{D} _\delta \tilde{\Gamma} ^\alpha _{\beta \gamma}+ \tilde{\Gamma} ^\nu_{\beta \delta} \tilde{\Gamma} ^\alpha _{\nu \gamma} -\tilde{\Gamma} ^\nu _{\beta \gamma} \tilde{\Gamma} ^\alpha _{\nu \delta}.
\end{equation}
The relation between the two Riemann tensors is given by:
\begin{eqnarray}
\tilde{R}^ \alpha _{\beta \gamma \delta} =R^ \alpha _{\beta \gamma \delta} \pm (1-q) \frac{\Delta}{R_0} \left[ \mu \right] \pm  \cdots
\end{eqnarray}
where $ \mu =\left[ \beta\partial_\gamma \Gamma^\alpha _{\beta \delta} + \partial_\gamma \tilde{\alpha}- \beta\partial_\delta \Gamma^\alpha _{\beta \gamma} - \partial_\delta \tilde{\alpha} \pm \tilde{\alpha}(\Gamma ^\nu_{\beta \delta} \pm \Gamma ^\alpha _{\nu \gamma} \mp \Gamma ^\nu _{\beta \gamma} \mp \Gamma ^\alpha _{\nu \delta}) \right]$. The Ricci tensor is
\begin{equation}
\tilde{R} _{\alpha \beta} = R_{\alpha \beta} \pm (1-q) \frac{\Delta}{R_0} \left[ \eta \right] \pm \cdots 
\end{equation}
Similarly Ricci scaler is : 
\begin{equation}
\tilde{R}=R \pm (1-q) \frac{\Delta}{R_0} \left[ \tau \right] \pm \cdots
\end{equation} 
where $\eta$ and $\tau $ are some constants. The Einstein field equations get modified as well:
\begin{equation}
\tilde{R} _{\alpha \beta}  - \frac{1}{2} g_{\alpha \beta} \tilde{R} = 8 \pi G T_{\alpha \beta} \pm (1-q) \frac{\Delta}{R_0}  \left [ \eta -  \tau \frac{1}{2} g_{\alpha \beta}  \right ] \pm \cdots.
\end{equation}
 So, the dynamics of the gravitational fields get modified for the fractional generalisation.

\end{document}